\documentclass[aps,pre,twocolumn,amsmath,amssymb,showpacs,amsfonts]{revtex4}
\usepackage{epsfig}
\usepackage{graphicx}
\usepackage{dcolumn}
\usepackage{bm}

\begin{document}

\title{The impact of hydrodynamic interactions on the preferential concentration of inertial particles in turbulence}

\author{Itzhak Fouxon}
\affiliation{Raymond and Beverly Sackler School of Physics and Astronomy,
Tel-Aviv University, Tel-Aviv 69978, Israel}
\begin{abstract}

We consider a dilute gas of inertial particles transported by the turbulent flow. Due to inertia the particles concentrate preferentially outside vortices. The pair-correlation function of the particles' concentration is known to obey at small separations a power-law with a negative exponent, if the hydrodynamic interactions between the particles are neglected. The divergence at zero separation is the signature of the random attractor asymptoted by the particles' trajectories at large times. However the hydrodynamic interactions produce a repulsion between the particles that is non-negligible at small separations. We introduce equations governing the repulsion and show it smoothens the singular attractor near the particles where the pair correlation function saturates. The effect is most essential at the Stokes number of order one, where the correlations decrease by a factor of a few.

%Heavy particles in turbulence lag behind the fluid rotation in vortices, which makes the particles cluster outside the vorticity
%regions. Thus inertia brings enhanced probability to find two particles in the flow close to each other. The resulting
%distribution of the inter-particle distance obeys a power-law with a negative exponent. We study the effect
%of the hydrodynamic interactions between the particles. These produce a repulsive force at small separations that produces a depletion
%of the probability of close particles. We introduce
%the equations governing the collisions of particles in turbulence with the account of the hydrodynamic interactions, and
%describe the depletion.

\end{abstract}
 \maketitle

The effect of inertia on the motion of particles driven by the stationary turbulent flow %of the ambient fluid
has enjoyed much attention of the researchers recently \cite{MaxeyRiley,Maxey,BFF,FFS1,Bec,FalkovichPumir,Collins,Stefano,Cencini,Olla,MehligWilkinson,IF}.
The progress in the understanding of the behavior of tracer particles \cite{review} made it natural to try to understand the impact of inertia that real particles have always. It was recognized quite early that inertia brings new qualitative effects \cite{Maxey} and
even small inertia can have a profound effect on the spatial distribution of particles in the flow \cite{FFS1}. However small the inertia is, the particles' motion in real space has a random attractor: at large times the trajectories asymptote a multi-fractal set that evolves in time
keeping its statistical properties constant \cite{FFS1,Bec}. The full set of fractal dimensions of the attractor was found recently \cite{IF,Tzhak} for real turbulence. The derivation involves no modeling and considers the statistics of turbulence as given but unknown. The statistics of density is log-normal and determined completely by the pair-correlation function \cite{Tzhak}. The latter obeys a power-law with a negative exponent at small separations \cite{BFF,FFS1,Bec,FalkovichPumir,Collins,Stefano,Cencini,Olla,MehligWilkinson,IF}.
The divergence at zero separation is the signature of the singular spatial structure of the attractor, while the exponent gives the correlation codimension of the attractor. These results are derived neglecting the hydrodynamic interactions, while one of the main applications is to collisions where the particles come near each other and the interactions become important. An example is the formation of rain in clouds where droplets can grow thanks to the collisions with other droplets, see \cite{FFS1,PK} and references therein. The rate of the collisions is proportional to the probability that turbulent fluctuations bring the particles together and it is measured by the pair-correlation function of the density at the separation equal to the diameter of the particles. The hydrodynamic interactions produce a repulsion that can be strong enough to prevent the collision completely. An example is the fall of a heavier particle onto the lighter one in the still air under the action of gravity. The repulsive force grows as the inverse of the distance between the particles' surfaces, and in the frame of hydrodynamics the particles never collide \cite{PK,Davis66,DS,HJ,Davis72}.

%It is therefore important to study the impact of hydrodynamic interactions on the particles collisions in turbulence.

In this Letter we derive equations that include the hydrodynamic interactions and describe the effect of the interactions on the pair-correlation function of particles' density. The analysis is performed at small inertia, as measured by the dimensionless Stokes number $St$. The depletion of the power-law in the pair-correlation function due to hydrodynamic interactions is found. The results are extrapolated to assess the impact of hydrodynamic interactions at $St\sim 1$, where the preferential concentration is maximal \cite{Cencini}.

%We use that the pair-correlation at a given separation is formed by events where two particles approach each other from large distances to this separation. We assume that during the approach from a cutoff separation $l_{cutoff}$ one may neglect the collisions with other particles, while the pair-correlation of density at the separation $l_{cutoff}$ is given by the squared mean density.
%This parallels the method used in the usual kinetic theory to find the two-particle distribution function of the dilute gas, where one
%tracks the particles back in time to the distance where they do not interact, and assumes that at that distance the two-particle distribution is given by the product of the single-particle distribution functions ("molecular chaos assumption") \cite{Landau}. While the interactions of inertial particles are indirect and mediated by the fluid, the analogy to the ordinary gas will be seen to be robust.

The hydrodynamic interactions between two particles do not admit exact analytic expressions and a simplification is necessary to perform the theoretical
analysis \cite{PK,Davis66,DS,HJ,Davis72,HB}. For heavy particles considered here the hydrodynamic interactions are determined by the perturbation flow caused by the inertial slip of the particles with respect to the surrounding flow. This flow obeys linear equations and the method of superposition
is available to derive the forces acting on the particles as sums of special cases.
%Representing the solution as the sum of special cases with particles moving along the line of centers and perpendicular to it, one finds that
The hydrodynamic interactions are important mainly when
%particular situation of
two particle move toward each other along the line of centers with equal velocities \cite{PK}. Say, when the distance between the surfaces of the particles measured in the units of their radius is $0.01$ the difference between the forces in this case and all other cases is about two orders of magnitude \cite{PK,Davis66}. To capture the main effect we account only for the hydrodynamic interactions caused by the motion of the particles toward each other. This is reasonable as this is the component of the motion describing the particles' approach.

We assume that without particles the forces stirring the fluid produce the turbulent flow $\bm u(t, \bm x)$ and the pressure $p_0(t, \bm x)$ that solve the Navier-Stokes (NS) equations \cite{Frisch}.
%The two-particle problem is set by considering the introduction of two spherical particles with equal radii $a$ in the fluid.
The flow in the presence of particles $\bm v(t, \bm x)$ obeys
\begin{eqnarray}&&\!\!\!\!\!\!
\partial_t\bm v+\bm v\cdot\nabla\bm v=-\nabla p +\nu\nabla^2\bm v,\ \ \nabla\cdot\bm v=0,  \label{NS}
\end{eqnarray}
where the particles are accounted via the no-slip boundary condition (b. c.) $\bm v(|\bm x-\bm x_i(t)|=d/2)=\bm v_i(t)$ and far from the particles the flow $\bm v$ and the pressure $p$ must match $\bm u(t, \bm x)$ and $p_0(t, \bm x)$. Here $\bm x_i$ and $\bm v_i$ are the particles' coordinates and velocities, respectively. The particles'  diameter $d$ is assumed to be much smaller that the Kolmogorov scale of turbulence $\eta$ determined by the kinematic viscosity $\nu$, see \cite{Frisch}. The flow around particles separated from the rest by distances much larger than $d$ is the sum of $\bm u$ and the Stokes flow caused by the relative motion of the particle through the flow at the speed $\bm v_i-\bm u(t, \bm x_i)$. For heavy particles the latter is damped via the linear friction force $\tau \dot {\bm v_i}=\bm u[t, \bm x_i(t)]-\bm v_i$
where $\tau$ is the Stokes' time and the Reynolds number $Re_p$ associated with the perturbation flow is assumed small \cite{MaxeyRiley}. The efficiency of relaxation is measured by the Stokes number $St\equiv \lambda\tau$, where $\lambda\sim \nu/\eta^2$ is the characteristic value of $|\nabla \bm u|$ and $\lambda^{-1}$ is the minimal time-scale of variations of $\bm u$. At $St\ll 1$ the particles follow the flow closely and one finds $\bm v_i=\bm u-\tau[\partial_t \bm u+(\bm u\cdot\nabla) \bm u]$, where the RHS is evaluated at $\bm x_i$. The estimate for $Re_p$ based on the latter expression validates $Re_p\ll 1$ at $St\ll 1$. Thus at $St \ll 1$ one can introduce the particles' velocity field
\begin{eqnarray}&&
\dot {\bm x_i}=\bm V[t, \bm x_i(t)],\ \ \bm V\equiv \bm u-\tau[\partial_t+\bm u\cdot\nabla]\bm u.\label{a2}
\end{eqnarray}
While $\bm V-\bm u$ is relatively small, it brings a qualitatively new effect: $\nabla\cdot \bm V=-\tau \nabla_j u_i\nabla_i u_j\neq 0$, though $\nabla\cdot \bm u=0$. The continuity equation $\partial_t n+\nabla\cdot(n\bm V)=0$ on the particles' density $n$ has no constant solution at $\tau>0$. The particles trajectories approach a multi-fractal set in space and the steady-state fluctuations of the density obey
\begin{eqnarray}&&\!\!\!\!\!\!\!\!\!\!\!\!
\langle n(0)n(\bm r)\rangle=(\eta/r)^{-\mu},\ \ r\ll \eta, \ \ \mu=2\left|\sum \lambda_i/\lambda_d\right|
%\propto St^2
,\label{a4}
\end{eqnarray}
where the angular brackets stand for spatial averaging \cite{BFF,FFS1,Bec,FalkovichPumir,Collins,Stefano,Cencini,Olla,MehligWilkinson,IF}, $\langle n\rangle=1$ and $\lambda_i$ are the Lyapunov exponents \cite{Oseledets} of the particles' flow $\bm V$. Both $\sum\lambda_i$ and $\lambda_d$ are negative and their leading order behavior at $St\ll 1$ is $|\sum\lambda_i|\propto St^2$ and $\lambda_d\approx \lambda_d(St=0)=\lambda_d^u$, where $\lambda_d^u\sim \lambda$ is the $d-$th Lyapunov exponent of $\bm u$. The latter determines the exponential divergence of fluid particles' in turbulence back in time \cite{FFS1,IF,review}. Importantly for the following, $\partial_t \bm u$ contribution into $\bm v-\bm u$ in Eq.~(\ref{a4}) has zero divergence and can be neglected, as can be seen explicitly from the complete solution \cite{IF}. Eq.~(\ref{a4}) is derived analytically \cite{FFS1} at $St\ll 1$. As $St$ increases, the power-law holds for $St<St_{critical}$ where $\mu$ first increases and then decreases to become zero at $St=St_{critical}$. At $St>St_{critical}$ the power-law breaks down \cite{Bec}. The preferential concentration is maximal \cite{Cencini} at $St\sim 1$, but significant \cite{Bec} already at $St\ll 1$. This is because $St\to 0$ is a singular limit: $\lim_{r\to 0}\lim_{St\to 0}\langle n(0)n(\bm r)\rangle=1$, while $\lim_{St\to 0}\lim_{r\to 0}\langle n(0)n(\bm r)\rangle=\infty$, so at a sufficiently small scale the inertia is always important. %Still at $St\ll 1$, by $\mu\ll 1$, there is a scale $l_{cut}\ll \eta$ such that $\langle n(0)n(|\bm r|=l_{cut})\approx 1$, cf. \cite{IF}.

The hydrodynamic interactions demand a different framework: the particle's velocity is determined not only by its position, but also by the position of the particle with which it interacts, so the particles' flow is no longer defined. To find $\langle n(0)n(\bm r)\rangle$ we use that it is proportional to the PDF $P(\bm R)$ of the distance $\bm R$ between two particles in the steady state. The existence of the latter demands a cutoff $R_{cut}$ at large $R$, as on average the two particles disperse. The cutoff does not enter the final result after the proportionality constant is fixed by the demand that $\langle n(0)n(\bm r)\rangle$ tends to $\langle n\rangle^2=1$ at large $r$. Say if the cutoff is due to a finite volume, then the particles spend most of the time at separations comparable with the volume's size. Occasionally they are brought close by the random flow. The frequency of these events is measured by $P(\bm R)$.

We first re-derive $\langle n(0)n(\bm R)\rangle$ from $P(R)$ at $R\gg d$ where the hydrodynamic interactions are negligible and one can use Eq.~(\ref{a2}). At $St\ll 1$ or $\mu\ll 1$ there are no fluctuations of $n$ at $R\sim \eta$, see Eq.~(\ref{a4}), so we assume $R\ll \eta$. Then
the distance obeys $\dot {\bm R}=\bm V[t, \bm x(t)+\bm R(t)]-\bm V[t, \bm x(t)]\approx \sigma \bm R$. Here $\sigma_{ij}=\nabla_jV_i[t, \bm x=\bm x(t)]$ can be expressed via $s_{ij}=\nabla_ju_i[t, \bm x=\bm x(t)]$ as $\sigma=s-\tau[{\dot s}+s^2]$, see Eq.~(\ref{a2}). By incompressibility
$tr s=0$, so $tr \sigma=-tr s^2$. One has $\bm R(t)=W(t)\bm R(0)$, where ${\dot W}=\sigma W$, so
\begin{eqnarray}&&\!\!\!\!\!\!\!\!\!\!\!\!\!
P(\bm R)\!\equiv\!\langle \delta\left[\bm R(t)\!-\!\bm R\right]\rangle\!=\!\left\langle
%\exp[\tau\int_0^t tr s^2(t')dt']
\delta\left(\bm R(0)\!-\!W^{-1}\bm R\right)/\det W\right\rangle.\label{equation}
\end{eqnarray}
Introducing $\zeta=\tau tr s^2$ one may write $\det W(t)=\exp(-\int_0^t \zeta(t')dt')$. From the definition
$|\sum \lambda_i|\equiv -\lim_{t\to\infty}\ln \det W(t)/t$ we have $\langle\zeta\rangle=|\sum \lambda_i|$, see \cite{Oseledets}. The "Fluctuation-Dissipation Theorem" ("FDT") holds
$\langle \zeta \rangle=(1/2)\int \langle\langle \zeta(0)\zeta(t)\rangle\rangle dt$, where the double angular brackets designate dispersion \cite{FF}.
%Note the statistics of $\nabla_ju_i$ in the particle's frame $\bm x(t)$, that defines $s_{ij}$, differs from the one of
%$\nabla_ju_i$ in the frame of fluid particles. For example, the latter obeys $\langle tr [\nabla_ju_i]^2\rangle=0$, as it is equal to the Eulerian average of the full derivative $\nabla\cdot [\bm u\cdot\nabla]\bm u$. In contrast, the inertial particles cluster outside vortices where $tr s^2>0$, and in the particle's frame $\langle tr s^2\rangle=|\sum\lambda_i|/\tau>0$.
%We now consider
At $t$ much larger than the correlation time $\lambda^{-1}$ of $\sigma$, see \cite{Frisch}, one can neglect in Eq.~(\ref{equation}) the contribution of the time interval of order $t_c$ near $t=0$ and perform independent averaging over $W(t)$ and the velocity field at $t<0$. The latter averaging averages $\bm R(0)$ producing the steady state condition %equation
%on $P(\bm R)$
%\begin{eqnarray}&&
$P(\bm R)= \left\langle \exp\left(\int_0^t \zeta(t')dt'\right)P\left(W^{-1}\bm R\right)\right\rangle$,
%\label{int}\end{eqnarray}
cf. \cite{BFL,FP}. The solution (independently of $R_{cut}$) is $P(\bm R)\propto R^{-\alpha}$,% where %$\alpha$ obeys
\begin{eqnarray}&&\!\!\!\!\!\!\!\!\!\!\!\!\!
\left\langle \!\exp\left(\int_0^t \zeta(t')dt'\right)\!\left|W^{-1}(t){\hat R}\right|^{-\alpha}\!\right\rangle\!=\!1,\ \
{\hat R}\!\equiv\! \bm R/R.\label{a6}
\end{eqnarray}
At $St\ll 1$ one has $\alpha\ll 1$ giving $|W^{-1}(t){\hat R}|^{-\alpha}\approx \exp[-\alpha|\lambda_d^u| t]$. Furthermore at $St\ll 1$ one can use Gaussian approximation for $\langle \exp\left(\int_0^t \zeta(t')dt'\right)\rangle$ which using the FDT gives
$\langle \exp\left(\int_0^t \zeta(t')dt'\right)\rangle\approx \exp\left(2t|\sum \lambda_i| \right)$. The substitution in Eq.~(\ref{a6}) reproduces $\alpha=\mu$ in Eq.~(\ref{a4}).

The observation that the divergenceless part of the inertial correction $\bm V-\bm u$ can be omitted from $\bm V$ without affecting the results to the leading order in $St$ implies
\begin{eqnarray}&&\!\!\!\!\!\!\!\!\!\!\!\!\!
\dot {\bm R}=\sigma'\bm R=s\bm R-\zeta \bm R/D,\ \ \sigma'_{ij}=s_{ij}-\delta_{ij}\zeta/D, \label{a8}
\end{eqnarray}
where $D=2, 3$ is the dimension, should give the same $\alpha$ (the part of $\sigma$ describing the divergence of $\bm V$ is $\zeta \delta_{ij}/D$). This can be verified substituting $W=W'$ in Eq.~(\ref{equation}) where $\dot{W'}=\sigma'W'$.
Equation (\ref{a8}) says that outside the vortices, at $tr s^2>0$, the particles are attracted due to inertia. Inertia produces a spherically symmetric contraction or expansion of all $\bm R$, impossible for incompressible flow, while the rest of its effects are negligible.

Using Eq.~(\ref{a6}) with $W=W'$ one also sees that the statistics of $s$ in Eq.~(\ref{a8}) can be considered as the one of $\nabla_j u_i$ in the fluid particle's frame and not the particle's frame, while $s$ can be assumed independent of $\zeta$. The values of $|W^{-1}{\hat R}|$ that determine the average in Eq.~(\ref{a6}) are $\exp(|\lambda_d^u|t)$, so $P(\bm R)$ is determined by the events for which $R$ decreases exponentially from the most probable, large, values of $R$ at the rate $|\lambda_d^u|$.
This is the same for fluid particles as $|\lambda_d^u|$ determines the rate of exponential divergence of trajectories backward in time. Inertia changes the volume of points that contract at this rate.

To account for hydrodynamic interactions, we consider two particles separated by $R\sim d$, well-separated from the rest. We look for the flow
around the particles in the form $\bm v(\bm x)=\bm u(x)+\bm U[\bm x-\bm x(t)]$ and $p(\bm x)=p_0(x)+P[\bm x-\bm x(t)]$. Substitution in Eq.~(\ref{NS}) and the use of $Re_p\ll 1$ give,
%the same assumptions as for the single-particle Stokes flow before,
\begin{eqnarray}&&
\nabla P=\nu\nabla^2\bm U,\ \ \bm U(x=d/2)=\dot {\bm x}-\bm u\left[t, \bm x(t)\right],\ \ \\&&
\bm U(|\bm x-\bm R|=d/2)=\dot {\bm x}+\dot {\bm R}-\bm u\left[t, \bm x(t)+\bm R(t)\right],
\end{eqnarray}
with $\bm U$, $P$ vanishing at large $\bm x$. This is the problem of two particles moving at speeds
$\bm J_1=\dot {\bm x}-\bm u\left[t, \bm x(t)\right]$ and $\bm J_2=\dot {\bm x}+\dot {\bm R}-\bm u\left[t, \bm x(t)+\bm R(t)\right]$ in the fluid at rest at infinity \cite{PK}. The motion can be written as ($i=1, 2$)
\begin{eqnarray}&&\!\!\!\!\!\!\!\!\!\!\!\!\!
\bm J_i=(-1)^iJ_r{\hat R}/2+\bm J_i-(-1)^iJ_r{\hat R}/2,\ \ \bm J\equiv \bm J_2-\bm J_1%i=1, 2
. \label{s1}
%\\&&\bm Y_2=\frac{\left[\bm W\cdot {\hat R}\right]{\hat R}}{2}+\bm Y_2-\frac{\left[\bm W\cdot {\hat R}\right]{\hat R}}{2}.\label{s2}
\end{eqnarray}
where $J_r\equiv \bm J\!\cdot\! {\hat R}$.
We consider the forces $\bm F_i$ on the particles as the superposition of the forces produced by the first term and the rest. At $J_r<0$ the former describes particles' motion toward each other along the line of centers at the same speed $J_r/2$. In this case the particles experience equal resistance
$F(R)|J_r|/2\tau$ (we set the mass to unity). At $R\gg d$ Stokes' law holds, $F(R)\approx 1$. For $R\sim d$ the flow is not the superposition of two Stokes flows and $F(R)\sim 1/(1-d/R)$, see \cite{PK}. This behavior makes the collisions in the still air impossible \cite{PK}.
We use $F(R)=1/(1-d/R)$ for qualitative estimates. The rest of the terms in Eq.~(\ref{s1}) do not change the distance between the particles and the deviation from Stokes' law for them is much weaker \cite{PK}. Studying the impact of hydrodynamic interactions on the particles' approach to each other, we account for the main effect and neglect the latter deviations%for the modification of the Stokes force only for forces arising from the first term in Eq.~(\ref{s1})
, so that
% Then the forces $\bm F_i$ on the particles are
%\begin{eqnarray}&&\!\!\!\!\!\!\!\!\!\!\!\!\!
$\tau\bm F_i=-\bm J_i-(-1)^i\theta\left(-J_r\right)J_r\left[F(R)-1\right]{\hat R}/2$,
%\nonumber\\&&
%\frac{d\bm v_2}{dt}=-\frac{1}{\tau}\left[\bm U_2-\frac{\left[\bm U\cdot {\hat R}\right]{\hat R}}{2}\right]
%-\frac{F(R)}{\tau}\frac{\left[\bm U\cdot {\hat R}\right]{\hat R}}{2}\nonumber
%\end{eqnarray}
where $\theta(x)$ is the step function. Subtracting the equations on $\ddot{\bm x}=\bm F_1$ and $\ddot{\bm x}+\ddot{\bm R}=\bm F_2$ and considering
$R\ll \eta$ we find
\begin{eqnarray}&&\!\!\!\!\!\!\!\!\!
\tau \dot {\bm w}\!=\!s\bm R\!-\!\bm w\!+\!
\left[1\!-\!F(R)\right]\left(w_r\!-\!{\hat R}s\bm R\right)
\theta\left[{\hat R}s {\bm R}\!-\!w_r\right]{\hat R},\nonumber
\end{eqnarray}
where $\bm w\equiv \dot{\bm R}$. At $St=0$ one finds the equation for fluid particles ${\bm w}=s\bm R$: there are no interactions for inertia-less particles that cause no friction. Solving the equation to the first order in $St$ and using $\sigma=s-\tau[{\dot s}+s^2]$ we find
\begin{eqnarray}&&\!\!\!\!\!\!\!\!\!
\dot {\bm R}=\sigma\bm R+z\theta\left(z\right)\left[1-F^{-1}(R)\right]\bm R,\ \ z\equiv \tau {\hat R}[{\dot s}+s^2]{\hat R}.\nonumber
\end{eqnarray}
To understand the effect of the hydrodynamic interaction force, which is radial, we multiply with ${\hat R}$. Introducing $\rho=\ln(R/d)$ and $\xi={\hat R}s{\hat R}$ we have ${\dot \rho}=\xi-z/F(R)$ at $z>0$.  Compared with  ${\dot \rho}=\xi-z$ holding without interactions, the hydrodynamic interactions deplete the inertial correction to the radial component of the relative velocity by the factor of $F(R)$, when the correction describes approaching particles. Thus the interactions are expected to eliminate the inertial enhancement of $P(R)$ at $R\sim d$, so $P(\bm R)$ and $\langle n(0)n(\bm R)\rangle$ saturate at $R\sim d$.

The effective description explained previously accounting only for the trace of ${\dot s}+s^2$ gives ($F^{-1}=1-d/R$)
\begin{eqnarray}&&
\dot {\bm R}=s\bm R-[1-d/R]\zeta\bm R/D,\label{a10}
\end{eqnarray}
for $\zeta>0$ and Eq.~(\ref{a8}) otherwise. Equation (\ref{a10}) transforms to Eq.~(\ref{a8}) at $R\gg d$ and thus reproduces the results at $R\gg d$, while capturing the main effect of the hydrodynamic interactions at $R\sim d$: the cancelation of the approaching inertial component of the relative velocity. The particles can collide due to $s\bm R$ term but the inertial enhancement of the collision velocity vanishes. Equation (\ref{a10}) should be supplied with the boundary conditions (b. c.) at $R=d$, describing the physics of the collision, e. g. absorbing b. c. for
coalescence. Consider the events that form the inertial enhancement of $P(R)$ at $R\sim d$ and thus have $tr s^2>0$. As one tracks $R(t)$ back in time it grows exponentially at the rate $|\lambda_d|$. The interactions become negligible at
$t\sim -|\lambda_d|^{-1}$ where $R(t)\gg d$. Since $tr s^2$ varies over the same time-scale $|\lambda_d|^{-1}$, then the behavior of $P(R)$ at $R\sim d$ should be captured by assuming $tr s^2>0$ during all the time-interval the interactions are relevant. Thus to understand the behavior of $P(R)$ at $R\sim d$ we use Eq.~(\ref{a10}) for any sign of $\zeta$. We find
\begin{eqnarray}&&\!\!\!\!\!\!\!\!\!\!\!\!\!
\dot{\rho}=\xi-\left(1-\exp[-\rho]\right)\zeta/D,\ \ \dot {{\hat R}}=s{\hat R}-\xi{\hat R}.\label{angle}
\end{eqnarray}
The dynamics of the orientation decouples and $\xi$ in the equation on $\rho$ can be considered as given, cf. \cite{FP}.
As mentioned, $s$ (and thus $\xi$) and $\zeta$ in can be considered independent where the statistics of $s_{ij}$ can be considered as the one of $\nabla_ju_i$ in the fluid particle's frame. Then $\xi$ is the same noise that governs the exponential separation of fluid particles, so e. g. $\langle \xi\rangle$ is the first Lyapunov exponent of the motion of fluid particles, cf. \cite{BFL,FP,review}. The Kraichnan model \cite{review} prescribes $\xi=\lambda_1^u+\xi'$ where $\xi'$ is a white noise $\langle \xi'(t_1)\xi'(t_2)\rangle=2\lambda_1^u\delta(t_2-t_1)/D$. This and $\zeta=|\sum \lambda_i|+\zeta'$ define the Kraichnan model for the problem, where $\zeta'$ is a white noise which amplitude is fixed by the "FDT" as $\langle \zeta'(t_1)\zeta'(t_2)\rangle=2|\sum \lambda_i|\delta(t_2-t_1)$.
One can write the Fokker-Planck equation on the PDF $P(\rho)$ of $\rho$ which gives that in the steady state the probability current must be constant. The constant is zero for reflecting b. c. at $\rho=0$. This gives the condition
\begin{eqnarray}&&\!\!\!\!\!\!\!\!\!\!\!\!\!
\frac{P'}{P}=\frac{2D^2-\mu D+\mu(D-1)\exp[-\rho]+\mu\exp[-2\rho]}
{2D+\mu\left(1-\exp[-\rho]\right)^2},\nonumber
%\\&&\!\!\!\!\!\!\!\!\!\!\!\!\!
%\approx D-\mu+\frac{\mu(3D-1)}{2D}\exp[-\rho]-\frac{\mu(D-1)}{2D}\exp[-2\rho]\nonumber
\end{eqnarray}
where $\mu=2|\sum \lambda_i|/|\lambda_d|$, see Eq.~(\ref{a4}), and we used \cite{review} that in the Kraichnan model $\lambda_1=|\lambda_d|$. Expanding the RHS to the first order in $\mu\ll 1$ and integrating,
\begin{eqnarray}&&\!\!\!\!\!\!\!\!\!\!\!\!\!
P=N\exp\Bigl(\rho D-\mu\rho+\mu(3D-1)\left(1-\exp[-\rho]\right)/(2D)
\nonumber\\&&
-\mu(D-1)\left(1-\exp[-2\rho]\right)/(4D)
\Bigr),\label{a11}
\end{eqnarray}
where $N$ is the normalization factor. %determined by $R_{cut}$. 
At $\mu=0$ we recover $P(\rho)\propto \exp[\rho D]$ equivalent to $P(\bm R)=const$. The hydrodynamic interactions produce the last two terms in the exponent. At $\exp[-\rho]\ll 1$ or $r\gg d$ these saturate at a constant and $\langle n(0)n(\bm r)\rangle\propto P[\ln(r/d)](d/r)^D$ reproduces  Eq.~(\ref{a4}). At $\rho\ll 1$ the linear term in the Taylor expansion of the hydrodynamic interactions' terms cancels $-\mu \rho$ and $\langle n(0)n(\bm r)\rangle\approx {\tilde N}\exp[\mu(D+1)\ln^2(R/d)/(4D)]$ depends on $r$ weakly. Between $\rho\ll 1$ and $\exp[-\rho]\ll 1$ there is a smooth transition between the two behaviors.  These modifications are less important at $\mu\ll 1$ where $(\eta/10 d)^{\mu}\approx (\eta/d)^{\mu}$ and $(\eta/r)^{\mu}$ near the particles is almost the same as far from them. In contrast, at $St\sim 1$ where Eq.~(\ref{a4}) holds with $\mu\sim 1$, the effect is important and $\langle n(0)n(d)\rangle$ is lowered by the factor of a few, see Eq.~(\ref{a11}).

\begin{figure}
\includegraphics[width=7.9 cm,clip=]{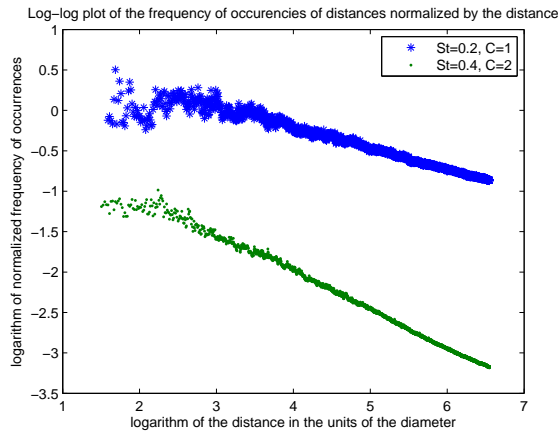}
\caption{Shown is the plot of $\log \left(P(\rho)\exp[-2\rho]\right)$ versus $\rho$. The power-law starts saturating at $\rho\approx 2.1$ for $St=0.4$ and $\rho\approx 2.4$ for $St=0.2$.}
\label{Figure}
\end{figure}

The above features are expected to be generic. We verify this numerically for a model where $\sigma$ has a correlation time of order of $\lambda^{-1}$, mimicking turbulence.
We consider the $D=2$ random renewal model \cite{FP} where the traceless, zero mean matrix $s_{\alpha\beta}(t)$ is a piecewise constant process, $s(t)=s^k$ at $k\Delta t\leq t\leq (k+1)\Delta t$ %independent random matrices
generated by
%\begin{eqnarray}&&\!\!\!\!\!\!\!
$s^k_{\alpha\beta}=C(\Delta  t)^{-1}
\left[\epsilon_{\alpha\beta 3}V^k+f^k_{\alpha}f^k_{\beta}/\sqrt{2}-\delta_{\alpha\beta}(f^k)^2/2\sqrt{2}\right]$,
%\nonumber\end{eqnarray}
where $\epsilon_{\alpha\beta\gamma}$ is the antisymmetric symbol,
$V^k$ and $\bm f^k_{\alpha}$ are independent standard Gaussian variables. 
%(having zero mean and unit dispersion)
%$\langle V^k V^l\rangle=\delta^{kl}$ and $\langle f^k_{\alpha}f^l_{\beta}\rangle=\delta^{kl}\delta_{\alpha\beta}$.
At $C\to 0$ one gets the Kraichnan model \cite{FP}. We consider $C\sim 1$.
%the correlation time $\Delta t$ is of order $\lambda^{-1}$ (say $\lambda_1^u\Delta t\approx 0.75$ at $C=2$). 
We model $\zeta$ as $|\sum\lambda_i|+\zeta'$ where the zero-mean Gaussian noise $\zeta'$ is renewed each $\Delta t$ and obeys $|\sum\lambda_i|=\langle \zeta^2\rangle\Delta t/2$ to ensure the "FDT". The results of $4\times 10^6$ renewals for $C=2$, $St=0.4$ and 
$C=1$, $St=0.2$, where $St^2\equiv |\sum\lambda_i|\Delta t$ are shown in Fig.~\ref{Figure}. When $\rho$ reached $\rho=0$ it was reset at $\rho_{cut}=1800$ (underestimating $P(\rho)$ near $\rho=0$). At
$\rho=\rho_{max}$ reflecting b. c. were used. The results agree qualitatively with the Kraichnan model.

We introduced the equations describing the two-particle dynamics of inertial particles at small $St$ with the account of hydrodynamic
interactions. The interactions eliminate the inertial effects in the particles' vicinity and the divergent power-law in the correlation function of the density saturates. The density is smooth at scales $r\sim d$, in contrast to the singular density in the single-particle approximation. Our equations are useful even without the interactions, for problems such as collisions of water droplets in clouds mediated by turbulence. 
%where the absorbing b. c. allow for the droplets' merging.
 
%There the equations must be used with a mixture of reflecting and  allowing for the possibility the droplets merge at the collision.

This work was supported by the ISF Grant No. $671/09$ and the BSF Grant No. $2010314$.

\end{document}